\documentclass[submission]{eptcs}
\providecommand{\event}{GandALF 2015} 
\usepackage{breakurl} 

\usepackage{epsfig,epic,eepic}
\usepackage{graphicx}
\usepackage{booktabs}
\usepackage{subfig}
\usepackage{hyperref}
\usepackage{import}
\usepackage{amsmath}
\usepackage{amstext}
\usepackage{amsfonts}
\usepackage{amssymb}
\usepackage[normalem]{ulem}
\usepackage{paralist}
\usepackage[percent]{overpic}
\usepackage{placeins}

\usepackage[printonlyused,withpage]{acronym}

\usepackage{stmaryrd} 

\usepackage{times}
\usepackage[table]{xcolor}
\usepackage{xspace}

\usepackage[noend]{algorithmic}

\usepackage[noline,scleft,noend,linesnumbered]{algorithm2e}
\SetInd{0.6em}{1em}

\usepackage{afterpage, longtable, footnote, framed, latexsym} 

\long\def\comment#1{}

\definecolor{Blue}{rgb}{0.25,0.33,0.77}
\definecolor{Red}{rgb}{1,0,0}
\definecolor{Green}{rgb}{0,1,0}

\newcommand{\mySmallHeading}[1]{\par\smallskip\noindent\textbf{#1}.~}

\newcommand{\R}{\mathbb{R}} 
\newcommand{\Rpos}{\ensuremath{\mathbb{R}^{\geq 0}}} 
\newcommand{\N}{\ensuremath{\mathbb{N}}} 
\newcommand{\Npos}{\ensuremath{\mathbb{N}^+}} 

\newcommand{\fun}[1]{{\textsl{#1}}} 

\newcommand{\wrt}{with respect to\xspace}

\def\fail{\textsl{FAIL}\xspace} 
\def\pass{\textsl{PASS}\xspace} 

\newcommand{\Reals}{\ensuremath{\mathbb{R}}\xspace} 
\newcommand{\RealsGEZ}{\ensuremath{\Reals^{\geq 0}}\xspace} 
\newcommand{\RealsGZ}{\ensuremath{\Reals^{+}}\xspace} 

\newcommand{\Rgez}{{\Reals^{\geq 0}}} 

\acrodef{DES}{Discrete Event System}
\acrodef{DG}{Disturbance Generator}
\acrodef{RDTGP}[DGP]{Random Disturbance Trace Generation Process}
\acrodef{EMC}{Explicit Model Checking}
\acrodef{HILS}[HILS]{Hardware In the Loop Simulation}
\acrodef{MDES}{Monitored \ac{DES}}
\acrodef{OEP}[OP]{Omission Probability} 

\acrodef{SLV}{System Level Verification}
\acrodef{SUV}{System Under Verification}

\acrodef{USG}{Uniform Random Scenario Generation}

\acrodef{SLFV}[SLFV]{System Level Formal Verification}

\acrodef{LBT}{Labels Branching Tree}
\acrodef{ZDB}{Zero Disturbance Block}

\acrodef{ROP}[RSG]{Random Sequence Generator}

\newtheorem{definition}{Definition}
\newtheorem{remark}{Remark}
\newtheorem{example}{Example}
\newtheorem{theorem}{Theorem}
\newtheorem{lemma}{Lemma}

\title{Simulator Semantics for System Level Formal Verification}

\author{Toni Mancini \qquad Federico Mari \qquad  Annalisa Massini \qquad  Igor Melatti \qquad  Enrico Tronci
\institute{
Computer Science Department - 
Sapienza University of Rome\\
Via Salaria 113, I-00198 Roma, Italy}
\email{\{tmancini,mari,massini,melatti,tronci\}@di.uniroma1.it} 
}
\def\titlerunning{Simulator Semantics for Model Checking Driven Simulation}
\def\authorrunning{T. Mancini, F. Mari, A. Massini, I. Melatti \&  E. Tronci}

\begin{document}
\maketitle



\begin{abstract}
%

Many simulation based \emph{Bounded Model Checking} approaches to \ac{SLFV} have been devised. Typically such approaches exploit the capability of simulators to save computation time by saving and restoring the state of the system under simulation. However, even though such approaches aim to (bounded) formal 
verification, as a matter of fact, the simulator behaviour is not formally modelled and the proof of correctness of the proposed approaches basically relies on the intuitive notion of simulator behaviour.  This gap makes it hard to check if the optimisations introduced to speed up the simulation do not actually omit checking relevant behaviours of the system under verification.

The aim of this paper is to fill the above gap by presenting a formal semantics for simulators.

\end{abstract}

\section{Introduction}
\label{intro.tex}

\acf{SLV} of Cyber-physical Systems has the goal of verifying that the \emph{whole} (i.e., software + hardware) system meets the given specifications.
\acf{HILS} is currently the main workhorse for system level verification and is supported by \emph{Model Based Design} tools such as Simulink,
(\url{http://www.mathworks.com})
VisSim,
(\url{http://www.vissim.com})
and Modelica
(\url{https://www.modelica.org/}).
In \ac{HILS} the \emph{actual software} reads [sends] values from [to] mathematical models (\emph{simulation}) of the physical systems  (e.g. engines, analog circuits, etc.) it will be interacting with. 


Our \ac{SUV} state can take discrete values (e.g., from the software state) as well as continuous values (e.g., from the physical system state).
Thus our \ac{SUV} can be conveniently modelled as a \emph{Hybrid System} (e.g., see \cite{Alur-emsoft-2011} and citations thereof) whose inputs belong to a finite set of uncontrollable events (\emph{disturbances}) modelling failures in sensors or actuators,  variations in the system parameters, etc.

We focus on \emph{deterministic systems} (the typical case for control systems), and model nondeterministic behaviours (such as faults) with disturbances.
Accordingly, in our framework, a \emph{simulation scenario} is just a finite sequence of disturbances.

A system is expected to \emph{withstand} all disturbance sequences that may arise in its 
operational scenarios. Correctness of a system is thus defined with respect to such \emph{admissible} disturbance sequences and the goal of \ac{HILS} is exactly that of showing that indeed the considered  \ac{SUV} can withstand all admissible disturbance sequences.
The set of admissible disturbance sequences typically satisfies constraints like the following:
1) the number of failures occurring within a certain period of time is less than a given threshold;
2) the time interval between two consecutive failures is greater than a given threshold;
3) a failure is repaired within a certain time, etc.

We focus on \ac{HILS} based \emph{Bounded} \ac{SLFV} of \emph{safety} properties.
That is, given a time horizon $T$ and a time step $\tau$ (time quantum between disturbances) our \ac{HILS} campaign returns \pass if there is no \emph{admissible} disturbance sequence of length $T$ and time step $\tau$ that violates the property under verification and \fail, along with a counterexample, otherwise.
In other words, \emph{Bounded} \ac{SLFV} is an \emph{exhaustive} (with respect to the set of admissible 
disturbance sequences) \ac{HILS} campaign.
In such a framework our exhaustive \ac{HILS} campaign works as a \emph{black box} \emph{bounded model checker} where the \ac{SUV} behaviour is defined by a simulator.


\subsection{Motivations  \label{motivations.tex} }

In our context the number of admissible disturbance sequences is finite since the number of disturbances is finite and the time horizon as well as the time quantum
between disturbances are both bounded.
Nevertheless the number of admissible disturbance sequences can be quite large. 
As a result, depending on the system considered and on the degree of assurance
sought a \ac{HILS} campaign may easily require months of simulation activity.

To decrease such a simulation time many \ac{HILS} based
\emph{Bounded Model Checking} approaches to \ac{SLFV} have been devised. 
Typically such approaches save simulation time by avoiding simulating more than once the 
same sequence of disturbances. 
This, in turn, is attained by exploiting the capability of simulators to save and restore a  simulation state.


However, even though such approaches aim to (bounded) formal verification, as a matter of fact the simulator behaviour is never formally modelled and the proof of correctness of the proposed approaches basically relies on the intuitive notion of simulator behaviour.  This gap makes it hard to check if the optimisations introduced to speed up the simulation do not actually omit checking of relevant behaviours of the system under verification.

The aim of this paper is to fill the above gap by presenting a \emph{formal} semantics for simulators and by proving \emph{soundness} and \emph{completeness} properties for it.





\subsection{Main Contributions 
\label{main-contributions}
}

Our main contributions can be summarised as follows.

We give a formal notion of simulator, of simulation campaign,
and provide a formal \emph{operational semantics} for simulators.

We show \emph{soundness} of our simulator semantics by showing that 
any simulation campaign defines a set of (\emph{in silico}) experiments 
that can be carried out on our \ac{SUV}.

We show \emph{completeness} of our simulator semantics by showing that 
any set of (\emph{in silico}) experiments to be carried out on our \ac{SUV} 
can be defined with a simulation campaign.

%

\subsection{Related Work}
 

\ac{SLFV} of cyber-physical systems via \ac{HILS} based bounded model checking has been studied in many
contexts. Here are a few examples.
Formal verification of Simulink models has been investigated in \cite{TSCC05,MBR06,WCMKS07}  
focusing on discrete time models (e.g., Stateflow or Simulink restricted to discrete time operators)
with small domain variables.
Formal verification of fully general Simulink models has been investigated in \cite{cav2013,dsd2014,pdp2014,pdp2015}. 
Formal verification of satellite operational procedures using ESA SIMSAT 
simulator has been investigated in \cite{SpaceOps2012}.

Simulation based approaches to statistical model checking have been widely investigated.
Here are just a few examples:  Simulink models for cyber-physical systems 
have been studied in \cite{fmsd2013}, mixed-analog circuits have been analyzed in
\cite{fmsd2010}, smart grid control policies have been considered in \cite{SmartGridComm2014}, and
biological models have been studied in \cite{bcb2013,fmcad2014,iwbbio2015}.


Of course \emph{Model Based Testing} (e.g., see \cite{model-based-testing-2005})
has widely considered automatic generation of test cases from models.
In our \ac{HILS} setting, automatic generation of simulation scenarios 
(for Simulink) has been investigated, for example, in
\cite{GYSRMS08,KAIRSS09,BHMK09,VSDP12}.    

Finally, synergies between simulation and formal methods have been widely investigated also in digital hardware verification.
Examples are in~\cite{%
validation-with-guided-search-dac98,%
smart-simulation-iccad00,%
guiding-simulation-dac06,%
abstration-guided-simulation-dac07}
and citations thereof.

All simulation based verification approaches considered in the literature heavily rely on carefully driving simulators in order to effectively carry out the planned verification activity.
However, to the best of our knowledge, none of them addresses the issue of providing a simulator semantics accounting for the simulator commands enabling saving and restoring of simulation states 
(the main simulation commands used to save simulation time).


\subsection{Outline of the paper}

Section~\ref{sec:DynSys} describes how we model disturbances as 
uncontrollable inputs to our (cyber-physical) \ac{SUV} that, in turn,
is modelled as a \emph{discrete event system}.
Section~\ref{sec:simulator} formalises the notion of simulator, 
simulation campaign and simulator semantics.
Section~\ref{sec:soundness} and Section~\ref{sec:completeness} provide, respectively, soundness and completeness theorems for our simulator semantics.

\section{Dynamical Systems}
\label{sec:DynSys}

\newcommand\RealOrderedPair{\ensuremath{(\RealsGEZ \times \RealsGEZ)^*}\xspace}

In this section we give the formal background on which our approach rests.
To this end, we model disturbances (Definition~\ref{discrete_sequence.def}) and define
our system (Definition~\ref{des.def}), by rephrasing the definition of dynamical system (see, e.g., \cite{sontag:book}).
Then, we define the simulation scenario, that is the sequence of disturbances occurring when the system starts from a given state, and the set of transitions associated to it.


Throughout the paper, we denote $\N$ the set of natural numbers, $\Npos$ the set of positive natural numbers,  \RealsGZ, \RealsGEZ  and \Reals the sets of  positive, non-negative and all real numbers, respectively.
Throughout the paper, we use \RealsGEZ to represent time and \RealsGZ to represent non-zero time durations.

\begin{figure}[t]
\begin{center}
\begin{overpic}[scale=0.8]{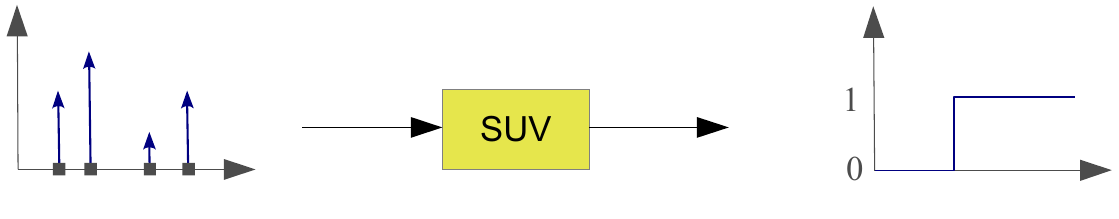}
\put(-6,15){\footnotesize{$\fun{u}(t)$}}
\put(19,0){\footnotesize{$t$}}
\put(31,10){\footnotesize{$\fun{u}(t)$}}
\put(54,10){\footnotesize{$\psi(\fun{u}, t)$}}
\put(66,15){\footnotesize{$\psi(\fun{u}, t)$}}
\put(96,0){\footnotesize{$t$}}
\put(9,0){(a)}
\put(45,0){(b)}
\put(86,0){(c)}
\end{overpic}
\caption{(a)~a discrete event sequence $\fun{u} \in {\cal U}^{\Rgez}$; (b)~our \ac{SUV}; (c)~the \ac{SUV} output $\psi(\fun{u},t)$ 
\label{fig:suv}}
\end{center}
\end{figure}

A \emph{discrete event sequence} (Definition~\ref{discrete_sequence.def} and Figure~\ref{fig:suv}(a)),
is a function associating to each (continuous) time instant a disturbance event
(such as a fault, a variation in a system parameter, etc).
We are considering a bounded time horizon, accordingly we require that the number of disturbances is finite, since no system can withstand an infinite number of disturbances within a finite time.
We represent with the real number 0 the event carrying no disturbance and with nonzero reals actual disturbances. 

\def\discreteSeqDef{\ensuremath{\fun{u}: \Rpos \to [0...d]}\xspace}
\def\eventListDef{\ensuremath{(\tau_0, u_0), (\tau_1, u_1), \ldots}\xspace}
\def\sqseq{$\llbracket e \rrbracket$}

\begin{definition}[Discrete event sequence] 
\label{discrete_sequence.def}
Let ${\cal U}$ be a finite subset of \Reals  such that $0 \in {\cal U}$. 
A \emph{discrete event sequence} over ${\cal U}$ is a function 
$u:\RealsGEZ \rightarrow {\cal U}$ such that the set 
$\{ t \in \RealsGEZ ~|~ \ u(t)\neq 0 \}$ has finite cardinality. 
We denote with ${\cal U}^{\Rgez}$ the set of discrete event sequences over ${\cal U}$.
We call \emph{time horizon} of a discrete event sequence $u$ the value $\max \{ t \in \RealsGEZ ~|~ \ u(t)\neq 0 \}$.
\end{definition}

An \emph{event list} provides an explicit representation for a discrete event sequence 
by listing pairs $(\tau,e)$, where $e>0$ is an event and $\tau$ is the time elapsed since the last (nonzero) event.

\begin{example} [Discrete event sequence and event list] 
\label{discrete_sequence.ex}
As a first example, let us consider the function $u$ defined as follows:
$$
u(t) = \begin{cases} 
1 & \mbox{for }\ \ t = k \tau, \ \mbox{where }\ k=1, 2, 3  \mbox{ and }\ \tau=1\\
0 & \mbox{otherwise} \end{cases} 
$$

The corresponding event list is: 
$[(\tau, 1), (\tau, 1), (\tau, 1)]$, and the time horizon of $u$ is $3 \tau$.
\end{example}

\begin{remark}
\label{u-expr.rem}
Let $\delta(t)$, with $t \in \R$, be the function such that if $t=0$ then $\delta(t)=1$, else  $\delta(t)=0$, that is  $\delta(t)$ represents a discrete impulse.
Any discrete event sequence $u(t)$ with horizon $h$ can be written in a unique way as finite sum  of discrete impulses, that is:
$$
u(t)= p_0 \delta(t) + \sum_{i=1}^h p_i \delta(t- \sum_{k=0}^{i-1} \tau_k)
$$
where $p_0 \in {\cal U}$, $p_i \in ({\cal U} - \{0\})$, and $\tau_k \in \RealsGZ$.
\end{remark}

\begin{example} [Impulses] 

Let us consider the discrete event sequence $u$ 
represented in Figure~\ref{fig:suv}(a).
It can be defined using impulses as follows:

$$
u(t)= 
2\delta(t- 3)+3\delta(t- 5)+ \delta(t- 10)+ 2\delta(t- 13)
$$

The event list associated to $u$ is: 
$[(3\tau, 2), (2\tau, 3), (5\tau, 1)$, $(3\tau, 2)]$.

\end{example}

Definition~\ref{des.def} formalizes how we model our \ac{SUV}, whereas Definition~\ref{restriction.def} and ~\ref{concatenation.def} define, respectively, the subset of ${\cal U}^{\Rgez}$ obtained as a restriction to a real interval, and the concatenation of two such subsets.

\begin{definition}
\label{restriction.def}
Let ${\cal U}^{\Rgez}$ be the set of discrete event sequences over the set ${\cal U}$. 
Given a discrete event sequence $u\in{\cal U}^{\Rgez}$ and two positive real numbers $t_1\leq t_2$, 
we denote with $u\mid_{[t_1,t_2)}$ the restriction of $u$ to the
interval $[t_1,t_2)$, i.e. the function
$u\mid_{[t_1,t_2)}:[t_1,t_2)\rightarrow {\cal U}$, 
such that $u\mid_{[t_1,t_2)}(t)=u(t)$ for all $t\in[t_1,t_2)$.
We denote ${\cal U}^{[t_1,t_2)}$ the \emph{restriction} of ${\cal U}^{\Rgez}$ to the domain $[t_1,t_2)$.
\end{definition}

\begin{definition}
\label{concatenation.def}
Assume that $t_1, t_2, t_3\in \RealsGEZ$ such that $t_1 < t_2 < t_3$.
If $\omega\in{\cal U}^{[t_1, t_2)}$ and $\omega'\in{\cal U}^{[ t_2,t_3)}$, their \emph{concatenation}, denoted as $\omega\omega'$, is the function 
$\tilde{\omega}\in{\cal U}^{[t_1, t_3)}$ defined as:
$$
\tilde{\omega}(t) = \begin{cases} 
\omega(t) &\mbox{if } t\in[t_1, t_2) \\
\omega'(t) & \mbox{if } t\in[ t_2,t_3) \end{cases} 
$$

\end{definition}

In our setting the system to be verified can be modelled as a continuous time \emph{Input-State-Output} deterministic dynamical system (see e.g.~\cite{sontag:book}) whose input functions are discrete event sequences, whose state can undertake continuous as well as discrete changes, and whose output ranges on any combination of discrete and continuous values.

\sloppy
\begin{definition}[Discrete Event System]
\label{des.def}
A \emph{Discrete Event System}, or simply DES, ${\cal H}$ is a tuple $({\cal X},{\cal U}, {\cal Y}, \varphi, \psi)$, where:
\begin{itemize}
\item ${\cal X}$, the \emph{state space} of ${\cal H}$, is a non-empty set whose elements denote \emph{states};
\item ${\cal U}$, the {\em input value space} of ${\cal H}$, is a finite subset of \Reals such that $0 \in {\cal U}$;
\item ${\cal Y}$, the \emph{output value space} of ${\cal H}$, is a non-empty set of \emph{outputs};
\item $\varphi: \RealsGZ \times {\cal X} \times {\cal U}^{\Rgez} \rightarrow {\cal X}$ 
is the \emph{transition map} of ${\cal H}$. 
Function $\varphi$ must satisfy the following properties:
\begin{itemize}
\item \emph{semigroup}: for each $t_1, t_2, t_3\in \RealsGEZ$ such that $t_1 < t_2 < t_3$, 
	$\omega\in{\cal U}^{[t_1, t_2)}$, 
	$\omega'\in{\cal U}^{[ t_2,t_3)}$, $x\in{\cal X}$ we have that 
	$\omega\omega'\in{\cal U}^{[t_1,t_3)}$ is such that 
	$\varphi(t_3-t_1, x, \omega\omega')=\varphi(t_3-t_2, \varphi(t_2-t_1, x, \omega), \omega')$; 
	
\smallskip

\item \emph{consistency}: for each $u \in {\cal U}$, $x \in {\cal X}$, we have $\varphi(0, x, u)=x$; 
	
\end{itemize}  
\item $\psi : \RealsGEZ \times {\cal X} \rightarrow {\cal Y}$ is the \emph{observation function} of ${\cal H}$. 

\end{itemize}
\end{definition}

Note that any simulator driven by its script language can be seen as a discrete event system.
This is why we focus on DES.

Our approach can model both the case in which the input is controllable, for example by control software (Example~\ref{pend.example}),  and the case in which the input is uncontrollable, for example disturbances such as faults (Examples~\ref{PendOnCart.example} and ~\ref{FCS.example}).

\begin{example}[Inverted Pendulum]
\label{pend.example}
A simple system is given by the Inverted Pendulum with Stationary Pivot Point, see e.g.~\cite{Kreisselmeier94,CDC12}.
The system is modeled by taking the angle $\theta$ and the angular velocity $\dot{\theta}$ as state variables. 
The input of the system is the torquing force $u$, that can influence the velocity in both directions. 
Moreover, the behaviour of the system depends on the pendulum mass $m$, the length of the pendulum $l$ and the gravitational acceleration $g$. 
Given such parameters, the motion of the system is described by the differential equation 
$\ddot{\theta} = \frac{g}{l} sin \theta + \frac{1}{ml^2} u$.

Let  ${\cal U}$ be $\{-1, 0, 1\}$, $\tau=10^{-6}$. Our discrete event system ${\cal H}$ is the tuple $({\cal X}, {\cal U}, {\cal Y}, \varphi, \psi)$, where:
\begin{itemize}
\item ${\cal X}= \R^2$ and ${\cal Y}=\R^2$;
\item $\varphi$ is solution to the system of differential equations:

$\dot{x_1}=x_2$

$\dot{x_2} = \frac{g}{l} sin x_1 + \frac{1}{ml^2} u$

where $x_1$ is the angle $\theta$ and $x_2$ is the angular velocity $\dot{\theta}$;
\item $\psi(t)$ is given by $[x_1(t),x_2(t)]$.

\end{itemize}

In Figure~\ref{fig:pendulum} the Simulink model of the inverted pendulum is shown, where we assume the pendulum mass $m=1$ and the length of the pendulum $l=1$.
Also we assume the function $u$ is given to the model as a sequence of values in the set $\{-1, 0, 1\}$.

\begin{figure}[h]
\begin{center}
\includegraphics[width=4.6in]{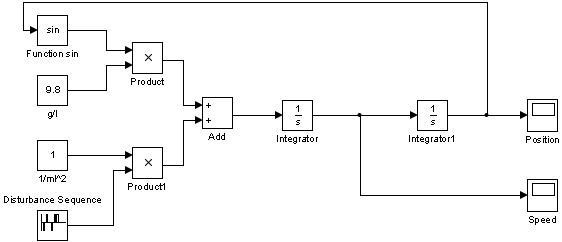}
\caption{Simulink model of the inverted pendulum.} 
\label{fig:pendulum}
\end{center}
\end{figure}

\end{example}

\begin{example}[Inverted pendulum on cart]
\label{PendOnCart.example}
Another example is given by the inverted pendulum on cart. For this system, the control input is the force $F$ that moves the cart horizontally and the outputs are the angular position of the pendulum $\theta$ and the horizontal position of the cart $x$.
The physical constraint between the cart and pendulum gives that both the cart and the pendulum have one degree of freedom each ($x$ and $\theta$, respectively). 
The  controlled system (the \emph{plant}) consists of the cart and the pendulum, whereas  the controller consists of the control software computing $F$ from the plant outputs ($x$ and $\theta$).
The dynamics of the system is described in the example available in the Simulink distribution. 
The Simulink model of the pendulum on cart, where disturbances are added, is shown in Figure~\ref{PendOnCart.fig}.

The system state is a pair $(z,w)$ where $z$ is the state of the control software and $w$ is the plant state, namely $w=[w_1, w_2, w_3,w_4]$, where:
$w_1=$ cart position, $w_2=$cart velocity , $w_3=$ pendulum angle, $w_4=$ pendulum angular velocity.

We model irregularities in the cart rail with a disturbance on the cart weight with respect to its nominal value 0.455 kg.
Let  ${\cal U}= \{0, 1, 2\}$ be our set of disturbances (see Definition \ref{discrete_sequence.def})  
modelling the fact that the cart weight is $(d+0.455)$, with $d\in {\cal U}$.
 
\begin{figure}[h]
\centering
\subfloat[Inverted pendulum on cart of the Simulink distribution]{%
\includegraphics[width=.7\linewidth]{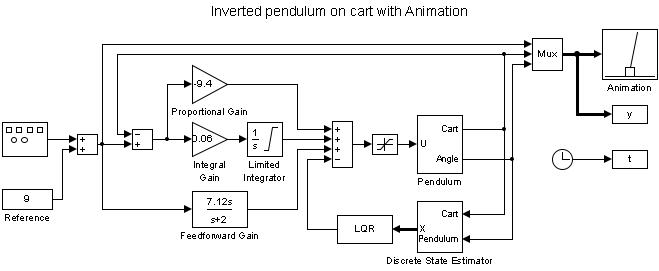}
\label{PendOnCart:subfig1}
}

\subfloat[Cart model with disturbances]{%
\includegraphics[width=.7\linewidth]{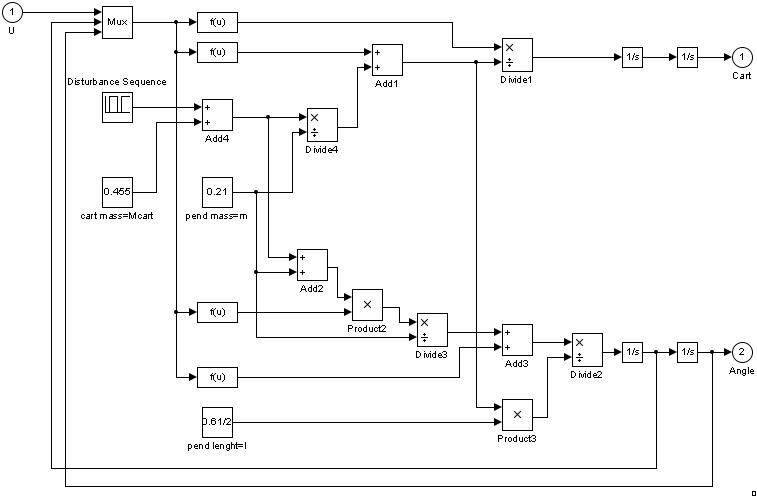}
\label{PendOnCart:subfig2}
}
\caption{Simulink model of the pendulum on cart with disturbances.}
\label{PendOnCart.fig}
\end{figure}

\end{example}

We will use the Inverted pendulum on cart
(Example~\ref{PendOnCart.example}) as running example throughout the paper.

\begin{example}[Fuel Control System]
\label{FCS.example}
The Fuel Control System (FCS) model in the Simulink distribution (see Figure~\ref{fig:FCSys})
has been studied in \cite{ClarkeHSCC10} using statistical model checking techniques, whereas the formal verification has been discussed in \cite{cav2013,pdp2014}.
The model is equipped with four sensors: throttle angle, speed, Oxygen in Exhaust Gas (EGO) and Manifold Absolute Pressure (MAP). 
In this case, the  tuple $({\cal X}, {\cal U}, {\cal Y}, \varphi, \psi)$ representing the discrete event system is:
\begin{itemize}
\item ${\cal X}$ is the set of plant (i.e., the engine) states along with the control software states;
\item ${\cal Y}$ is the set of plant outputs monitored by the control software;
\item ${\cal U}$ is the set of disturbance sequences that can be obtained assuming that only sensors EGO and MAP can fail, giving rise to disturbances 1 and 2, respectively; the minimum time between faults is one second and all faults are transient, that is disturbance 1 models a fault on sensor EGO, followed by a repair within one second, and disturbance 2 models a fault on sensor MAP, followed by a repair within one second too;
\item $\varphi$ computes the dynamics of the system states;
\item $\psi(t)$ computes the system output from the present system state.

\end{itemize}

\begin{figure}[ht]
\begin{center}
\vspace{-0.5cm}
\includegraphics[width=5.1in]{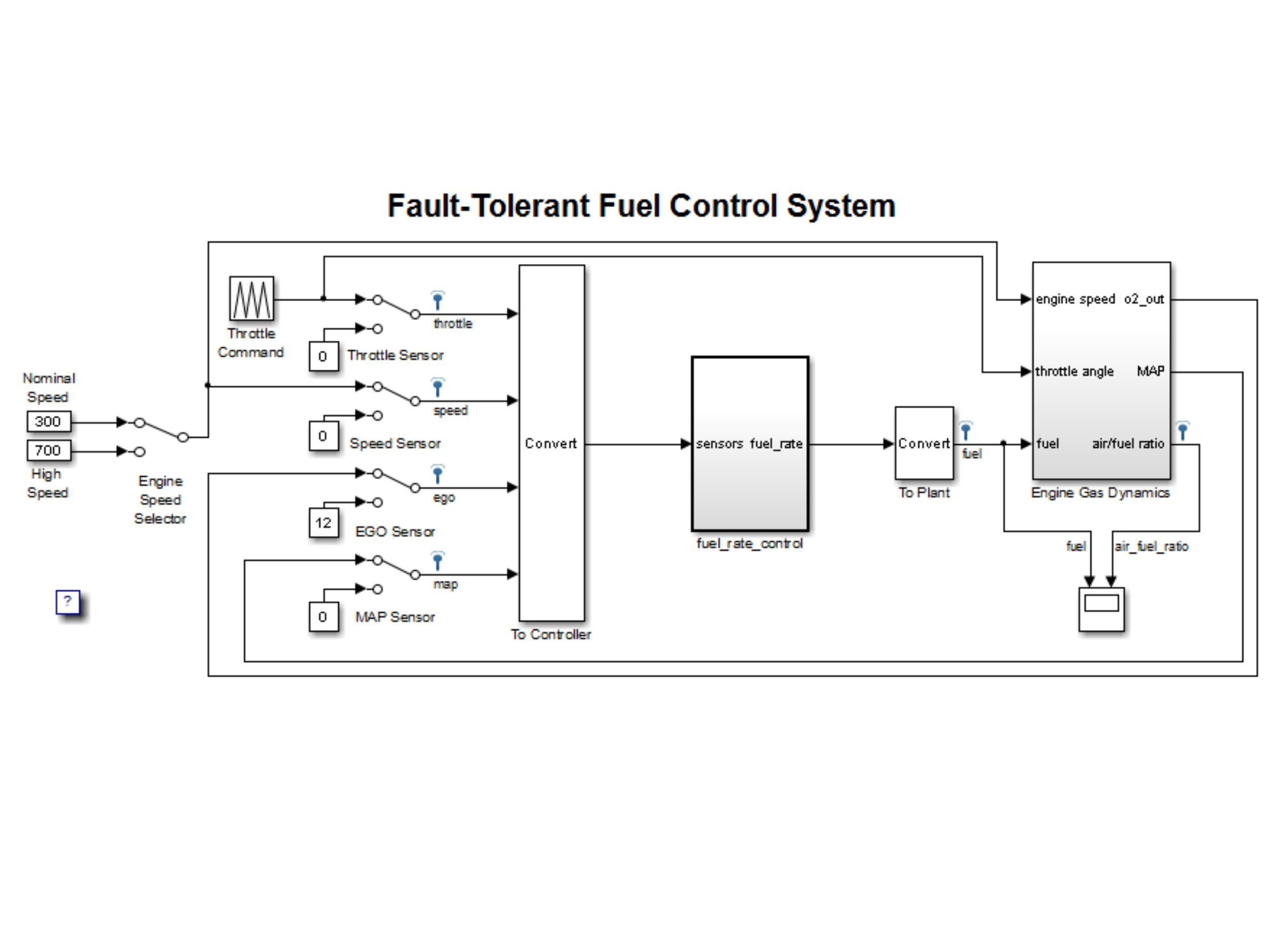}
\vspace{-2.5cm}
\caption{The Simulink Fuel Control System.}
\label{fig:FCSys}
\end{center}
\end{figure}


\end{example}

Our discrete event system as defined in Definition~\ref{des.def},  models a hybrid system describing a cyber physical system, as shown in Examples~\ref{pend.example}, \ref{PendOnCart.example}
 and \ref{FCS.example}.
For this reason, we denote our system with ${\cal H}$.

In the following we define the notion of simulation scenario, that is the sequence of disturbances received by our system  starting from a given initial state, and we give an example.

\begin{definition}[Simulation scenario]
A \emph{simulation scenario} for ${\cal H}$ is a pair $(x,u)$ where $x \in {\cal X}$ and $u \in {\cal U}^{\Rgez}$.
\end{definition}

\begin{example}[Simulation scenario]
\label{SimScenario.example}
Let  ${\cal H}$ be the Inverted pendulum on cart system described in Example~\ref{PendOnCart.example}.
Let $u(t)$ be the discrete event sequence defined as: $u(t)= \delta(t-0.04) + \delta(t-0.08)$ and let the initial state be $x_0=(z_0,[0,0,0,0])$, where $z_0$ is the control software initial state. Then, a simulation scenario for   ${\cal H}$ is $(x_0,u)$.

\end{example}

Definitions~\ref{trace.def} and \ref{transitionSet.def} give the definition of sequence of transitions and set of transitions explored by a \ac{SUV} under a given simulation scenario, respectively.

\begin{definition}[Trace of a simulation scenario]
\label{trace.def}
Let ${\cal H}$ be a DES and let $x \in {\cal X}$ be a state and $u \in {\cal U}^{\Rgez}$, $u(t)= p_0 \delta(t) + \sum_{i=1}^h p_i \delta(t- \sum_{k=0}^{i-1} \tau_k)$,  a discrete event sequence giving a simulation scenario $(x,u)$ for ${\cal H}$.
The \emph{trace} of the simulation scenario $(x,u)$, denoted $Tr(x,u)$, is the finite sequence of transitions $Tr(x,u)=[(x_0, p_0, 0,x_1), (x_1, p_1, \tau_1,x_2), \ldots, (x_{h-1}, p_{h-1}, \tau_{h-1},x_h) ]$ such that $x_0=x$ and $x_{i+1} =\varphi(\tau_i,  x_i, p_i \delta(t))$. 
\end{definition}

\begin{example}[Trace of a simulation scenario]
\label{TraceSimScenario.example}
Let  ${\cal H}$ be the Inverted pendulum on cart system described in Example~\ref{PendOnCart.example} and let $(x_0,u)$ be the simulation scenario of Example~\ref{SimScenario.example}.

The trace of $(x_0,u)$ is $Tr(x_0,u)=[(x_0, 0, 0,x_1), (x_1, 1, 0.04,x_2), (x_2, 1, 0.04,x_3) ]$, 
where $x_0=(z,[0,0,0,0])$ and the $x_i$ values, $i=1,2,3$ are obtained by running the simulation with the Simulink model shown in Example~\ref{PendOnCart.example} and are:
$x_1=(z_1,[-0.017,-0.881,0.057,2.914])$, $x_2=(z_2,[-0.049,-0.694,0.167,2.451])$ and $x_3=(z_3,[-0.072,-0,431,0.253,1.878])$.

\end{example}

\begin{definition}[Set of transitions of a simulation scenario]
\label{transitionSet.def}
The set of transitions associated to a simulation scenario $(x,u)$ is the set:
$$
{\cal T}_{(x,u)}= \{(z,p,\tau,z') | (z,p,\tau,z') \in Tr(x,u) \}
$$
\end{definition}

\begin{example}[Set of transitions of a simulation scenario]
\label{TransitionSet.example}
Let us consider system ${\cal H}$, simulation scenario $(x_0,u)$, and trace $Tr(x_0,u)$ as in Example~\ref{SimScenario.example}.
The set of transitions associated to $(x_0,u)$ is simply the set
$ {\cal T}_{(x,u)}=\{(x_0, 0, 0,x_1), (x_1, 1, 0.04,x_2), (x_2, 1, 0.04,x_3) \}$,  
where $x_1$, $x_2$ and $x_3$ assume the values specified in Example~\ref{SimScenario.example}.

\end{example}

%

\section{Simulators and Simulation Campaigns}
\label{sec:simulator}
In this Section we formalise the notion of discrete event system simulator (Definition~\ref{simulator.def} and Definition~\ref{simulator.commands.def}), of simulation campaign (Definition~\ref{sim_campaign.def}) and of set of transitions of a simulation campaign (Definition~\ref{trans_sim_camp}).

In many cases it is necessary to consider a huge number of simulation scenarios for having an exhaustive \ac{HILS}.
The overall number of simulation steps can be prohibitively large if each scenario is simulated from the initial state of the (\ac{SUV}) simulator.
The definition of set of transitions of a simulation campaign (Definition~\ref{trans_sim_camp}) helps to individuate states necessary to complete the simulation campaign avoiding to repeat the same sequence of commands several times.

\begin{definition}[Discrete Event System Simulator]
\label{simulator.def}
A \ac{DES} simulator ${\cal S}$ is a tuple $({\cal H}, W)$, where ${\cal H}=({\cal X}, {\cal U}, {\cal Y}, \varphi, \psi)$ is a \ac{DES} and $W$ is a finite set whose elements are called \emph{simulator states}.
Each $w \in W$ is a pair $(x, M)$ where $x \in {\cal X}$ is a state of ${\cal H}$ and $M$ is a finite subset of ${\cal X}$ that models the content of the \emph{simulator memory}.
\end{definition}	
	
Unless otherwise stated, in the following ${\cal S}$ is a simulator for the DES ${\cal H}$ as in Definition~\ref{simulator.def}. 

Note that, at the beginning the simulator memory contains the initial state $x_0$ of ${\cal H}$.

The semantics of simulator commands we use to execute our simulation scenarios, and the transition function $\xi$ are given in Definition~\ref{simulator.commands.def}.

\begin{definition}[Simulator commands and transition function]	
\label{simulator.commands.def}
\label{simulator.transition.function.def}
Let ${\cal S}$ be a simulator. 
\begin{itemize}
\item The commands for ${\cal S}$ are: $\fun{load}(x)$, $\fun{store}$, $\fun{free}(x)$, $\fun{run}(p, t)$, where $x \in {\cal X}$ is a  state of ${\cal H}$, $t \in \R^+$ is a time duration, and $p \in {\cal U}$ is an event ($x, t, p$ are command arguments).

\item The \emph{transition function} $\xi$ of ${\cal S}$,  defines how the internal state of the simulator ${\cal S}$ changes upon execution of a command. Namely:
$\xi(x, M, \fun{cmd}(\fun{args})) = (x', M')$ when the simulator ${\cal S}$ moves from internal state $(x, M)$ to state $(x', M')$ upon processing command $\fun{cmd}$ with arguments \fun{args}.

For each $x \in {\cal X}$, function $\xi$ is defined as follows:
\begin{itemize}	
\renewcommand{\labelitemi}{\tiny{$\bullet$}}
\item if $x' \in M$ then $\xi(x, M, \fun{load}(x')) = (x', M)$ 
		
\item if $x' \in M$ then $\xi(x, M, \fun{free}(x')) = (x, M \setminus \{x'\})$ 

\item $\xi(x, M, \fun{store}) = (x, M \cup \{x\})$ 
		
\item $\xi(x, M, \fun{run}(p,\tau)) = (x', M)$, 
where $x' = \varphi(\tau, x, u)$, where $u(t) = p \delta(t)$.
\end{itemize}
\end{itemize}
\end{definition}

Given a sequence of simulation scenarios, we can build a sequence of commands, \emph{simulation campaign}, driving the simulator through such scenarios. We define the simulator \emph{output sequence} as the sequence of the \ac{SUV} outputs associated to the simulator states traversed by a simulation campaign.
Conversely, given a simulation campaign, we can compute the sequence of scenarios  simulated by it. These concepts are formalised in Definition~\ref{sim_campaign.def}.

\begin{definition}[Simulation campaign and sequence of simulator states]
\label{sim_campaign.def}
Let ${\cal S}$ be a simulator and let $\xi$ be its transition function.  
\begin{itemize}
\item A \emph{simulation campaign} for ${\cal S}$ is a triple $\Xi=(x,M,\chi)$ , where $x \in {\cal X}$, $M\subset {\cal X}$ and $\chi$ is a sequence  (possibly empty or infinite) of commands along with their arguments, $\chi = \fun{cmd}_0(\fun{args}_0), \fun{cmd}_1(\fun{args}_1), \ldots $.
A simulation campaign consisting of a finite sequence of commands is a \emph{finite} simulation campaign or a simulation campaign of finite length; the length of $\chi=\fun{cmd}_0(\fun{args}_0), \ldots, \fun{cmd}_{c-1}(\fun{args}_{c-1}) $ is $c$ and it is denoted by $|\chi|=c$.

\label{sequence_of_sim_states.def}
\item The \emph{sequence of simulator states} of ${\cal S}$ \wrt a simulation campaign $\Xi=(x_0,M_0,\chi)$ is the sequence $(x_0, M_0), (x_1, M_1), \ldots $, where for all $j$, $\xi(x_j, M_j, \fun{cmd}_j(\fun{args}_j)) = (x_{j+1}, M_{j+1})$.

We denote with $\chi(x_0, M_0, j)$ the $j$-th element of such a sequence, that is $\chi(x_0, M_0, j)=(x_j, M_j)$. In other words $\chi(x_0, M_0, j)$ is the simulator state after the execution of the $j$-th command.
\item The set of simulator states with respect to a simulation campaign $\chi$ is denoted $\chi(x_0, M_0)$, that is $\chi(x_0, M_0)= \{ \chi(x_0, M_0, j) | j= 0,1, \ldots, |\chi|-1 \}$.

\end{itemize}
\end{definition}

\begin{example}[Simulation campaign]
\label{SimCamp.example}
Let ${\cal H}$ be the Inverted pendulum on cart considered in Example~\ref{PendOnCart.example} and let $(x_0,u)$ be the simulation scenario considered in Example~\ref{SimScenario.example}, where $u(t)= \delta(t-0.04) + \delta(t-0.08)$.

The simulation campaign $\Xi$ obtained by using this simulation scenario is the triple $\Xi=(x_0,\{x_0\},\chi)$, where $\chi$ is the sequence  of commands $\chi = (\fun{run}(0,0.04), \fun{run}(1,0.04), \fun{run}(1,0.04)) $.

The sequence of simulator states with respect to $\Xi$ is:
$$
(x_0,\{x_0\})\xrightarrow{\fun{run}(0,0.04)}(x_1,\{x_0\})\xrightarrow{\fun{run}(1,0.04)}(x_2,\{x_0\})\xrightarrow{\fun{run}(0,0.04)}(x_3,\{x_0\}).
$$
State values are obtained by running the simulation.

\medskip

An example of a  more complex simulation campaign, \!$\Xi_1$, can be obtained by considering the sequence of simulation scenarios $((x_0,u), (x_3,u_1),(x_3,u_2),(x_0,u_3))$, where:

$u(t)= \delta(t-0.04) + \delta(t-0.08)$ 

$u_1(t)= \delta(t-0.04) + \delta(t-0.08)  + \delta(t-0.12)$

$u_2(t)= \delta(t-0.12)$

$u_3(t)= \delta(t-0.04)  + 2 \delta(t-0.12) + \delta(t-0.16)  + 2 \delta(t-0.24)$.

A graphical representation of simulation campaign \!$\Xi_1$ is shown in Figure~\ref{fig:treeSimCamp}, where we see that the discrete event sequences $u(t)$ and $u_3(t)$ are  applied when having state $x_0$, and sequences $u_1$ and $u_2$ are applied having state $x_3$.

\begin{figure}[h]
\begin{minipage}{0.4\textwidth}
\begin{center}
\includegraphics[width=2.8in]{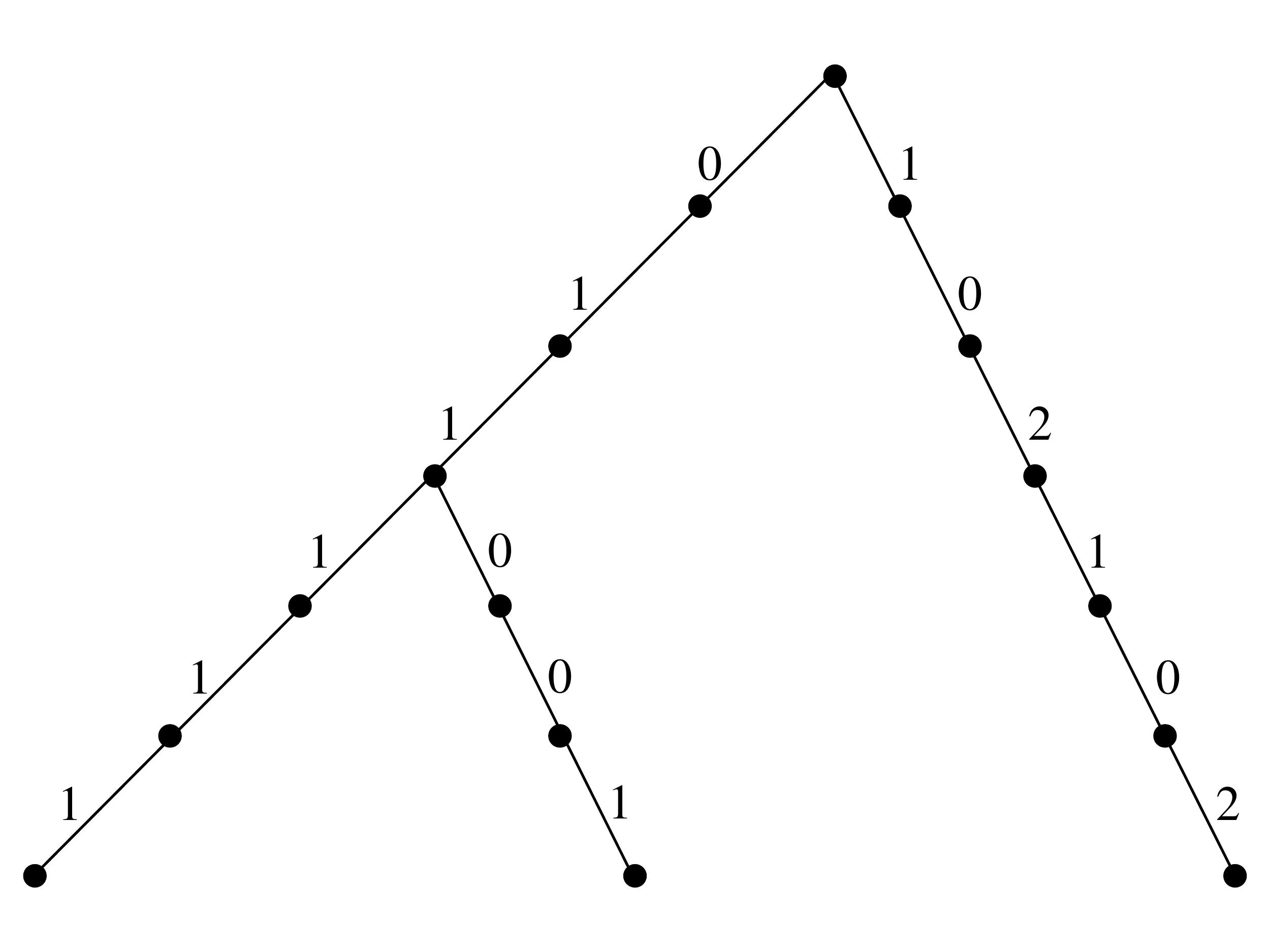}
\caption{A graphical representation of simulation campaign $\Xi_1$ (Example~\ref{SimCamp.example}).}
\label{fig:treeSimCamp}
\end{center}
\end{minipage}
\hspace{2cm}
\begin{minipage}{0.4\textwidth}
\begin{center}
\includegraphics[width=2.8in]{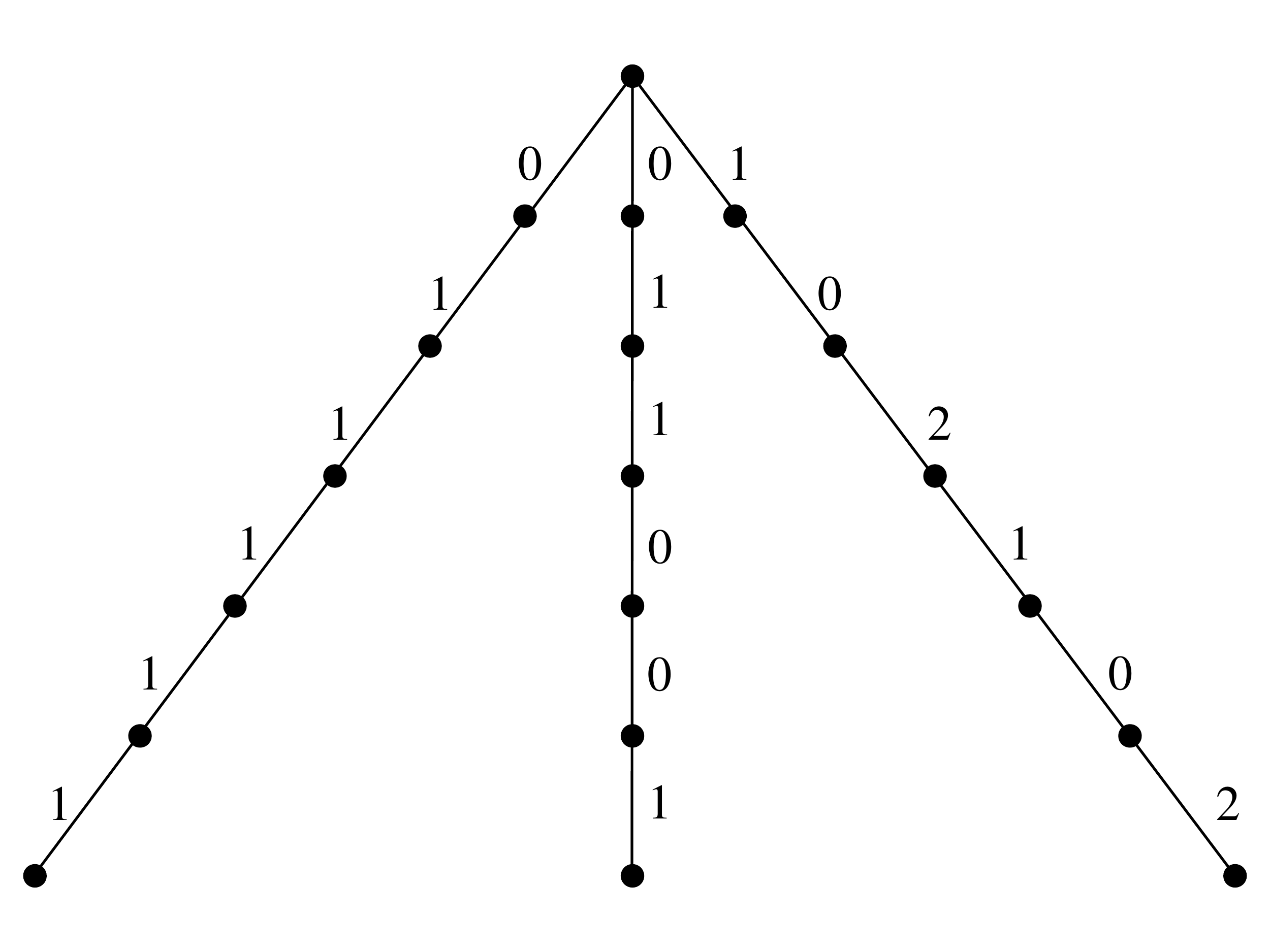}
\caption{A graphical representation of the normal simulation campaign $\Xi_2$ (Example~\ref{NormalSimCamp.example}).}
\label{fig:treeNormalSimCamp}
\end{center}
\end{minipage}
\end{figure}

The simulation campaign $\Xi_1$ obtained by using the sequence of simulation scenarios above is the triple $\Xi_1=(x_0,\{x_0\},\chi_1)$ , where $\chi_1$ is given by the following command sequence:

\begin{tabular}{lll}
$\chi_1=$& $\fun{run}(0,0.04)$, $\fun{run}(1,0.04)$, $\fun{run}(1,0.04)$, $\fun{store}$, \\
         & $\fun{run}(1,0.04)$, $\fun{run}(1,0.04)$, $\fun{run}(1,0.04)$, $\fun{load}(x_3)$,  \\
         & $\fun{run}(0,0.04)$, $\fun{run}(0,0.04)$, $\fun{run}(1,0.04)$, $\fun{free}(x_3)$, \\
				 & $\fun{load}(x_0)$, $\fun{run}(1,0.04)$, $\fun{run}(0,0.04)$, $\fun{run}(2,0.04)$, \\
			 	 & $\fun{run}(1,0.04)$, $\fun{run}(0,0.04)$, $\fun{run}(2,0.04)$\\
\end{tabular}  

\medskip


The sequence of simulator states with respect to $\Xi_1$ is:

$(x_0,\{x_0\})\xrightarrow{\fun{run}(0,0.04)}$
$(x_1,\{x_0\})\xrightarrow{\fun{run}(1,0.04)}$
$(x_2,\{x_0\})\xrightarrow{\fun{run}(1,0.04)}$
$(x_3,\{x_0\})\xrightarrow{\fun{store}}$
$(x_3,\{x_0,x_3\})\xrightarrow{\fun{run}(1,0.04)}$
$(x_4,\{x_0,x_3\})\xrightarrow{\fun{run}(1,0.04)}$
$(x_5,\{x_0,x_3\})\xrightarrow{\fun{run}(1,0.04)}$
$(x_6,\{x_0,x_3\})\xrightarrow{\fun{load}(x_3)}$
$(x_3,\{x_0,x_3\})\xrightarrow{\fun{run}(0,0.04)}$
$(x_7,\{x_0,x_3\})\xrightarrow{\fun{run}(0,0.04)}$
$(x_8,\{x_0,x_3\})\xrightarrow{\fun{run}(1,0.04)}$
$(x_9,\{x_0,x_3\})\xrightarrow{\fun{free}(x_3)}$
$(x_9,\{x_0\})\xrightarrow{\fun{load}(x_0)}$
$(x_0,\{x_0\})\xrightarrow{\fun{run}(1,0.04)}$
$(x_{10},\{x_0\})\xrightarrow{\fun{run}(0,0.04)}$
$(x_{11},\{x_0\})\xrightarrow{\fun{run}(2,0.04)}$
$(x_{12},\{x_0\})\xrightarrow{\fun{run}(1,0.04)}$
$(x_{13},\{x_0\})\xrightarrow{\fun{run}(0,0.04)}$
$(x_{14},\{x_0\})\xrightarrow{\fun{run}(2,0.04)}$
$(x_{15},\{x_0\})$
\end{example}


Definition~\ref{NormalSimCamp.def} gives the notion of normal simulation campaign, that is a simulation campaign for which every simulation scenario starts from an initial state.

\begin{definition}[Normal Simulation Campaign]
\label{NormalSimCamp.def}
A simulation campaign is in \emph{normal} form if it consists only of commands $\fun{load}$ and $\fun{run}$.
\end{definition}

\begin{example}[Normal Simulation Campaign]
\label{NormalSimCamp.example}
An example of simulation campaign in normal form is $\Xi_2=(x_0,\{x_0\},\chi_2)$, where $\chi_2$ is given by the following command sequence:

\begin{tabular}{lll}
$\chi_2=$& $\fun{run}(0,0.04)$, $\fun{run}(1,0.04)$, $\fun{run}(1,0.04)$, $\fun{run}(1,0.04)$, $\fun{run}(1,0.04)$, $\fun{run}(1,0.04)$,\\
         & $\fun{load}(x_0)$,  $\fun{run}(0,0.04)$, $\fun{run}(1,0.04)$, $\fun{run}(1,0.04)$, $\fun{run}(0,0.04)$, $\fun{run}(0,0.04)$, $\fun{run}(1,0.04)$,  \\
		 		 & $\fun{load}(x_0)$,  $\fun{run}(1,0.04)$, $\fun{run}(0,0.04)$, $\fun{run}(2,0.04)$, $\fun{run}(1,0.04)$, $\fun{run}(0,0.04)$, $\fun{run}(2,0.04)$\\
\end{tabular}  

\medskip

A graphical representation of $\Xi_2$ is shown in Figure~\ref{fig:treeNormalSimCamp}.


\end{example}

Note that, since a normal simulation campaign has no $\fun{store}$ commands, a command $\fun{load}$ can only load an initial state.

Definition~\ref{trans_sim_camp}, resting on Definition~\ref{sim_campaign.def}, defines the  set of transitions of a simulation campaign. 
Definition \ref{equiv_sim_camp.def} gives the notion of equivalent simulation campaigns.

\begin{definition}[Set of transitions of a Simulation Campaign]
\label{trans_sim_camp}
We denote with ${\cal T}_\Xi$ the set of transitions of ${\cal S}$ explored by $\Xi$, that is 
${\cal T}_\Xi= 
\{(x,p,\tau,x') | 
\exists\; M, M' \; 
[(x, M) \mbox{ is a simulator state of } {\cal S} \mbox{ wrt } \Xi 
\;\wedge\; 
\xi(x, M,\fun{run}(p,\tau)) = (x', M')]\}$.

\end{definition}

\begin{definition}[Equivalent simulation campaigns]
\label{equiv_sim_camp.def}
We say  that the simulation campaign $\Xi$ is equivalent to $\Xi'$ and we write $\Xi \sim \Xi'$ if ${\cal T}_\Xi={\cal T}_{\Xi'}$.
\end{definition}

\begin{example}[Equivalent simulation campaigns]
\label{EquivalentSimCamp.example}
The simulation campaign $\Xi_1$ in Example~\ref{SimCamp.example} and the normal simulation campaign $\Xi_2$ in Example~\ref{NormalSimCamp.example} are equivalent.

In fact the set of transitions explored by $\Xi_1$ and $\Xi_2$ is:

$
{\cal T}_{\Xi_1} = {\cal T}_{\Xi_2} = \{$
$(x_0,0,0.04,x_1)$,
$(x_1,1,0.04,x_2)$,
$(x_2,1,0.04,x_3)$,
$(x_3,1,0.04,x_4)$,
$(x_4,1,0.04,x_5)$,
$(x_5,1,0.04,x_6)$,
$(x_3,0,0.04,x_7)$,
$(x_7,0,0.04,x_8)$,
$(x_8,1,0.04,x_9)$,
$(x_0,1,0.04,x_{10})$,
$(x_{10},0,0.04,x_{11})$,
$(x_{11},2,0.04,x_{12})$,
$(x_{12},1,0.04,x_{13})$,
$(x_{13},0,0.04,x_{14})$,
$(x_{14},2,0.04,x_{15})\}$

This can also be easily seen looking at the set of edges (transitions) in Figures~\ref{fig:treeSimCamp} and \ref{fig:treeNormalSimCamp}.

\end{example}

Lemma~\ref{equiv.lemma} formalizes the fact that for each simulation campaign, we can determine an equivalent simulation campaign in which each simulation scenario starts from an initial state.

\begin{lemma}
\label{equiv.lemma}
Given  a simulation campaign $\Xi$ for a simulator ${\cal S}$, there exists a simulation campaign $\Xi'$ such that:
\begin{itemize}
\item $\Xi'$ is in normal form
\item $\Xi' \sim \Xi$
\end{itemize}
\end{lemma}

We give the idea of the proof by using the following example.

\begin{example}[Lemma~\ref{equiv.lemma}]
\label{lemma.example}
Consider the simulation campaign $\Xi_1$ in Example~\ref{SimCamp.example}.
$\Xi_1$ is not in normal form.
However, by modifying it so that all simulation scenarios (paths on the tree of Figure~\ref{fig:treeSimCamp}) start from the initial state, we get the normal simulation campaign $\Xi_2$ illustrated in Example~\ref{NormalSimCamp.example}.

Further, it follows from Example~\ref{NormalSimCamp.example} that $\Xi_1 \sim \Xi_2$.

\end{example}

\section{Soundness}
\label{sec:soundness}
In this section we show the soundness of our simulator semantics.
That is, we show that any simulation campaign stems from a set of simulation scenarios. 
This guarantees that any simulation campaign has indeed a physical (computational) meaning.

\begin{theorem}[Soundness]
\label{soundness.theo}
Given a simulation campaign $\Xi$ for ${\cal S}$ there exists a set ${\cal A}=\{ (x_1,u_1), (x_2,u_2), \ldots, (x_k,u_k)\}$ of simulation scenarios  such that
$$
{\cal T}_\Xi= \cup \{ {\cal T}_{(x_i,u_i)}| \;  i=1, \ldots, k \}.
$$
\end{theorem}

As we did for Lemma~\ref{equiv.lemma}, we give the idea of the proof by using an example.

\begin{example}[Soundness]
\label{soundness.example}
Consider the simulation campaign $\Xi_1$ in Example~\ref{SimCamp.example}.
The set of transitions of $\Xi_1$, ${\cal T}_{\Xi_1}$, is shown in Example~\ref{EquivalentSimCamp.example}. 

Now, let us consider the set ${\cal A}$ consisting of simulation scenarios $(x_0,u)$, $(x_3,u_1)$, $(x_3,u_2)$ and $(x_0,u_3)$, defined in Example~\ref{SimCamp.example}.

The sets of transitions for simulation scenario $(x_0,u)$ is shown in Example~\ref{TransitionSet.example}, and the sets of transitions associated to the other three simulation scenarios are, respectively:

$ {\cal T}_1$ = $ {\cal T}_{(x_3,u_1)}= \{$ $(x_3,1,0.04,x_4)$, $(x_4,1,0.04,x_5)$, $(x_5,1,0.04,x_6) \} $

$ {\cal T}_2$ = $ {\cal T}_{(x_3,u_2)}= \{$ $(x_3,0,0.04,x_7)$, $(x_7,0,0.04,x_8)$, $(x_8,1,0.04,x_9)\} $

$ {\cal T}_3$ = $ {\cal T}_{(x_0,u_3)}= \{$ $(x_0,1,0.04,x_{10})$, $(x_{10},0,0.04,x_{11})$, $(x_{11},2,0.04,x_{12})$, $(x_{12},1,0.04,x_{13})$, $(x_{13},0,0.04,x_{14})$, $(x_{14},2,0.04,x_{15})\}$

It is easy to see that ${\cal T}_{\Xi_1} = \cup \{ {\cal T}_j| \;  j=0, \ldots, 3 \}$.

\end{example}

%
%
%
%
%
%
%

\section{Completeness}
\label{sec:completeness}
In this section we show the completeness of our simulator semantics.
That is, we show that any set of simulation scenarios yields a simulation campaign. 
This guarantees that any set of physical experiments 
can be defined by a suitable simulation campaign.

\begin{theorem}
\label{completeness.theo}

Let ${\cal A}=\{ (x_1,u_1), (x_2,u_2), \ldots, (x_k,u_k)\}$ be a set of simulation scenarios of ${\cal H}$.
Then there exists a simulation campaign $\Xi$ for ${\cal S}$ such that
$$
{\cal T}_\Xi= \cup \{ {\cal T}_{(x_i,u_i)}| \; \; i=1, \ldots, k \}.
$$
\end{theorem}

Also for this theorem, we give the idea of the proof by using an example.

\begin{example}[Completeness]
\label{completeness.example}
Let us consider the set ${\cal A}$ consisting of simulation scenarios $(x_0,u)$, $(x_3,u_1)$, $(x_3,u_2)$ and $(x_0,u_3)$, defined in Example~\ref{SimCamp.example}.

The sets of transitions associated to these simulation scenarios, ${\cal T}_0= {\cal T}_{(x_0,u)}$, ${\cal T}_1={\cal T}_{(x_3,u_1)}$, ${\cal T}_2={\cal T}_{(x_3,u_2)}$ and ${\cal T}_3={\cal T}_{(x_0,u_3)}$, are shown in Example~\ref{soundness.example}.
Let $\bar{{\cal T}}$ be the set obtained as union of the sets of transitions above, that is $\bar{{\cal T}}= \cup \{ {\cal T}_j| \;  j=0, \ldots, 3 \}$.

Now, let us consider the simulation campaign $\Xi_1$ in Example~\ref{SimCamp.example} and the set of transitions of $\Xi_1$, ${\cal T}_{\Xi_1}$, shown in Example~\ref{EquivalentSimCamp.example}.

It is easy to see that $\bar{{\cal T}}={\cal T}_{\Xi_1}$.
\end{example}

%
%
%
%
%
%

\section{Conclusions}
\label{sec:conclusions}
\label{conclusions.tex}

We provided a formal notion of simulator, of simulation campaign.
and a formal \emph{operational semantics} for simulators.

Furthermore we showed \emph{soundness} and \emph{completeness}
of our simulator semantics by showing that 
\emph{any} 
simulation campaign defines a set of (\emph{in silico}) experiments 
for the \ac{SUV} (soundness)
and, conversely,
that \emph{any} such a set can be defined with a simulation campaign 
(completeness).

This work enables formal proofs of correctness 
for simulation based formal verification approaches and provides
formal tools enabling investigation of more aggressive approaches to the 
optimisation of the simulation activity entailed by \ac{HILS} based
 \ac{SLFV}.

\medskip\mySmallHeading{Acknowledgements}
Work partially supported by FP7 projects SmartHG 
(317761) and PAEON 
(600773).


\bibliographystyle{eptcs}
\bibliography{biblio_original,newbib}

\begin{thebibliography}{10}
\providecommand{\bibitemdeclare}[2]{}
\providecommand{\surnamestart}{}
\providecommand{\surnameend}{}
\providecommand{\urlprefix}{Available at }
\providecommand{\url}[1]{\texttt{#1}}
\providecommand{\href}[2]{\texttt{#2}}
\providecommand{\urlalt}[2]{\href{#1}{#2}}
\providecommand{\doi}[1]{doi:\urlalt{http://dx.doi.org/#1}{#1}}
\providecommand{\bibinfo}[2]{#2}

\bibitemdeclare{inproceedings}{CDC12}
\bibitem{CDC12}
\bibinfo{author}{V.~\surnamestart Alimguzhin\surnameend},
  \bibinfo{author}{F.~\surnamestart Mari\surnameend},
  \bibinfo{author}{I.~\surnamestart Melatti\surnameend},
  \bibinfo{author}{I.~\surnamestart Salvo\surnameend} \&
  \bibinfo{author}{\surnamestart E.Tronci\surnameend} (\bibinfo{year}{2012}):
  \emph{\bibinfo{title}{Automatic control software synthesis for quantized
  discrete time hybrid systems}}.
\newblock In: {\sl \bibinfo{booktitle}{Proc. 51th {IEEE} Conference on Decision
  and Control, {CDC}}}, \doi{10.1109/CDC.2012.6426260}.

\bibitemdeclare{inproceedings}{Alur-emsoft-2011}
\bibitem{Alur-emsoft-2011}
\bibinfo{author}{R.~\surnamestart Alur\surnameend} (\bibinfo{year}{2011}):
  \emph{\bibinfo{title}{Formal verification of hybrid systems}}.
\newblock In: {\sl \bibinfo{booktitle}{Proc. 11th Int. Conf. on Embedded
  Software, EMSOFT 2011, part of the Seventh Embedded Systems Week}},
  \bibinfo{publisher}{ACM}, \doi{10.1145/2038642.2038685}.

\bibitemdeclare{inproceedings}{BHMK09}
\bibitem{BHMK09}
\bibinfo{author}{A.~\surnamestart Brillout\surnameend},
  \bibinfo{author}{N.~\surnamestart He\surnameend},
  \bibinfo{author}{M.~\surnamestart Mazzucchi\surnameend},
  \bibinfo{author}{M.~Purandare \surnamestart D.~Kroening\surnameend},
  \bibinfo{author}{P.~\surnamestart R\"{u}mmer\surnameend} \&
  \bibinfo{author}{G.~\surnamestart Weissenbacher\surnameend}
  (\bibinfo{year}{2010}): \emph{\bibinfo{title}{Mutation-based Test Case
  Generation for Simulink Models}}.
\newblock In: {\sl \bibinfo{booktitle}{Proc. 8th Int. Conf. on Formal Methods
  for Components and Objects}}, \bibinfo{series}{FMCO'09},
  \bibinfo{publisher}{Springer-Verlag}, \doi{10.1007/978-3-642-17071-3}.

\bibitemdeclare{book}{model-based-testing-2005}
\bibitem{model-based-testing-2005}
\bibinfo{author}{M.~\surnamestart Broy\surnameend},
  \bibinfo{author}{B.~\surnamestart Jonsson\surnameend}, \bibinfo{author}{J.-P.
  \surnamestart Katoen\surnameend}, \bibinfo{author}{M.~\surnamestart
  Leucker\surnameend} \& \bibinfo{author}{A.~\surnamestart
  Pretschner\surnameend} (\bibinfo{year}{2005}):
  \emph{\bibinfo{title}{Model-Based Testing of Reactive Systems: Advanced
  Lectures}}.
\newblock {\sl \bibinfo{series}{LNCS}} \bibinfo{volume}{3472},
  \bibinfo{publisher}{Springer}, \doi{10.1007/b137241}.

\bibitemdeclare{article}{fmsd2010}
\bibitem{fmsd2010}
\bibinfo{author}{E.~M. \surnamestart Clarke\surnameend},
  \bibinfo{author}{A.~\surnamestart Donz{\'{e}}\surnameend} \&
  \bibinfo{author}{A.~\surnamestart Legay\surnameend} (\bibinfo{year}{2010}):
  \emph{\bibinfo{title}{On simulation-based probabilistic model checking of
  mixed-analog circuits}}.
\newblock {\sl \bibinfo{journal}{Formal Methods in System Design}}
  \bibinfo{volume}{36}(\bibinfo{number}{2}), \doi{10.1007/s10703-009-0076-y}.

\bibitemdeclare{inproceedings}{abstration-guided-simulation-dac07}
\bibitem{abstration-guided-simulation-dac07}
\bibinfo{author}{F.~M. \surnamestart De~Paula\surnameend} \&
  \bibinfo{author}{A.~J. \surnamestart Hu\surnameend} (\bibinfo{year}{2007}):
  \emph{\bibinfo{title}{An effective guidance strategy for abstraction-guided
  simulation}}.
\newblock In: {\sl \bibinfo{booktitle}{Proc. 44th annual Design Automation
  Conference}}, \bibinfo{series}{DAC '07}, \bibinfo{publisher}{ACM},
  \bibinfo{address}{New York, NY, USA}, \doi{10.1145/1278480.1278498}.

\bibitemdeclare{inproceedings}{SpaceOps2012}
\bibitem{SpaceOps2012}
\bibinfo{author}{Verzino \surnamestart G.\surnameend},
  \bibinfo{author}{F.~\surnamestart Cavaliere\surnameend},
  \bibinfo{author}{F.~\surnamestart Mari\surnameend},
  \bibinfo{author}{I.~\surnamestart Melatti\surnameend},
  \bibinfo{author}{G.~\surnamestart Minei\surnameend},
  \bibinfo{author}{I.~\surnamestart Salvo\surnameend},
  \bibinfo{author}{Y.~\surnamestart Yushtein\surnameend} \&
  \bibinfo{author}{E.~\surnamestart Tronci\surnameend} (\bibinfo{year}{2012}):
  \emph{\bibinfo{title}{Model checking driven simulation of sat procedures}}.
\newblock In: {\sl \bibinfo{booktitle}{Proc. of 12th International Conference
  on Space Operations (SpaceOps 2012)}}, \bibinfo{organization}{SpaceOps},
  \doi{10.2514/6.2012-1275611}.

\bibitemdeclare{inproceedings}{GYSRMS08}
\bibitem{GYSRMS08}
\bibinfo{author}{A.~A. \surnamestart Gadkari\surnameend},
  \bibinfo{author}{A.~\surnamestart Yeolekar\surnameend},
  \bibinfo{author}{J.~\surnamestart Suresh\surnameend},
  \bibinfo{author}{S.~\surnamestart Ramesh\surnameend},
  \bibinfo{author}{S.~\surnamestart Mohalik\surnameend} \&
  \bibinfo{author}{K.~C. \surnamestart Shashidhar\surnameend}
  (\bibinfo{year}{2008}): \emph{\bibinfo{title}{AutoMOTGen: Automatic Model
  Oriented Test Generator for Embedded Control Systems}}.
\newblock In: {\sl \bibinfo{booktitle}{Proc. 20th Int. Conf. Computer Aided
  Verification, CAV}}, \doi{10.1007/978-3-540-70545-1\_19}.

\bibitemdeclare{inproceedings}{smart-simulation-iccad00}
\bibitem{smart-simulation-iccad00}
\bibinfo{author}{P.~H. \surnamestart Ho\surnameend},
  \bibinfo{author}{T.~\surnamestart Shiple\surnameend},
  \bibinfo{author}{K.~\surnamestart Harer\surnameend},
  \bibinfo{author}{J.~\surnamestart Kukula\surnameend},
  \bibinfo{author}{R.~\surnamestart Damiano\surnameend},
  \bibinfo{author}{V.~\surnamestart Bertacco\surnameend},
  \bibinfo{author}{J.~\surnamestart Taylor\surnameend} \&
  \bibinfo{author}{J.~\surnamestart Long\surnameend} (\bibinfo{year}{2000}):
  \emph{\bibinfo{title}{Smart simulation using collaborative formal and
  simulation engines}}.
\newblock In: {\sl \bibinfo{booktitle}{Proc. 2000 IEEE/ACM Int. Conf. on
  Computer-aided design}}, \bibinfo{series}{ICCAD '00},
  \bibinfo{publisher}{IEEE Press}, \doi{10.1109/ICCAD.2000.896461}.

\bibitemdeclare{inproceedings}{KAIRSS09}
\bibitem{KAIRSS09}
\bibinfo{author}{A.~\surnamestart Kanade\surnameend},
  \bibinfo{author}{R.~\surnamestart Alur\surnameend},
  \bibinfo{author}{F.~\surnamestart Ivancic\surnameend},
  \bibinfo{author}{S.~\surnamestart Ramesh\surnameend},
  \bibinfo{author}{S.~\surnamestart Sankaranarayanan\surnameend} \&
  \bibinfo{author}{K.~C. \surnamestart Shashidhar\surnameend}
  (\bibinfo{year}{2009}): \emph{\bibinfo{title}{Generating and Analyzing
  Symbolic Traces of Simulink/Stateflow Models}}.
\newblock In: {\sl \bibinfo{booktitle}{Proc. 21st Int. Conf. Computer Aided
  Verification, CAV}}, \doi{10.1007/978-3-642-02658-4\_33}.

\bibitemdeclare{article}{Kreisselmeier94}
\bibitem{Kreisselmeier94}
\bibinfo{author}{G.~\surnamestart Kreisselmeier\surnameend} \&
  \bibinfo{author}{T.~\surnamestart Birkholzer\surnameend}
  (\bibinfo{year}{1994}): \emph{\bibinfo{title}{Numerical nonlinear regulator
  design}}.
\newblock {\sl \bibinfo{journal}{Automatic Control, IEEE Transactions on}}
  \bibinfo{volume}{39}(\bibinfo{number}{1}), \doi{10.1109/9.273337}.

\bibitemdeclare{inproceedings}{cav2013}
\bibitem{cav2013}
\bibinfo{author}{T.~\surnamestart Mancini\surnameend},
  \bibinfo{author}{F.~\surnamestart Mari\surnameend},
  \bibinfo{author}{A.~\surnamestart Massini\surnameend},
  \bibinfo{author}{I.~\surnamestart Melatti\surnameend},
  \bibinfo{author}{F.~\surnamestart Merli\surnameend} \&
  \bibinfo{author}{E.~\surnamestart Tronci\surnameend} (\bibinfo{year}{2013}):
  \emph{\bibinfo{title}{System Level Formal Verification via Model Checking
  Driven Simulation}}.
\newblock In: {\sl \bibinfo{booktitle}{Computer Aided Verification - 25th
  International Conference, {CAV}}}, \doi{10.1007/978-3-642-39799-8\_21}.

\bibitemdeclare{inproceedings}{dsd2014}
\bibitem{dsd2014}
\bibinfo{author}{T.~\surnamestart Mancini\surnameend},
  \bibinfo{author}{F.~\surnamestart Mari\surnameend},
  \bibinfo{author}{A.~\surnamestart Massini\surnameend},
  \bibinfo{author}{I.~\surnamestart Melatti\surnameend} \&
  \bibinfo{author}{E.~\surnamestart Tronci\surnameend} (\bibinfo{year}{2014}):
  \emph{\bibinfo{title}{Anytime System Level Verification via Random Exhaustive
  Hardware in the Loop Simulation}}.
\newblock In: {\sl \bibinfo{booktitle}{17th Euromicro Conference on Digital
  System Design, {DSD}}}, \doi{10.1109/DSD.2014.91}.

\bibitemdeclare{inproceedings}{pdp2014}
\bibitem{pdp2014}
\bibinfo{author}{T.~\surnamestart Mancini\surnameend},
  \bibinfo{author}{F.~\surnamestart Mari\surnameend},
  \bibinfo{author}{A.~\surnamestart Massini\surnameend},
  \bibinfo{author}{I.~\surnamestart Melatti\surnameend} \&
  \bibinfo{author}{E.~\surnamestart Tronci\surnameend} (\bibinfo{year}{2014}):
  \emph{\bibinfo{title}{System Level Formal Verification via Distributed
  Multi-core Hardware in the Loop Simulation}}.
\newblock In: {\sl \bibinfo{booktitle}{22nd Euromicro International Conference
  on Parallel, Distributed, and Network-Based Processing, {PDP}}},
  \doi{10.1109/PDP.2014.32}.

\bibitemdeclare{inproceedings}{pdp2015}
\bibitem{pdp2015}
\bibinfo{author}{T.~\surnamestart Mancini\surnameend},
  \bibinfo{author}{F.~\surnamestart Mari\surnameend},
  \bibinfo{author}{A.~\surnamestart Massini\surnameend},
  \bibinfo{author}{I.~\surnamestart Melatti\surnameend} \&
  \bibinfo{author}{E.~\surnamestart Tronci\surnameend} (\bibinfo{year}{2015}):
  \emph{\bibinfo{title}{SyLVaaS: System Level Formal Verification as a
  Service}}.
\newblock In: {\sl \bibinfo{booktitle}{23rd Euromicro International Conference
  on Parallel, Distributed, and Network-Based Processing, {PDP}}},
  \doi{10.1109/PDP.2015.119}.

\bibitemdeclare{inproceedings}{SmartGridComm2014}
\bibitem{SmartGridComm2014}
\bibinfo{author}{T.~\surnamestart Mancini\surnameend},
  \bibinfo{author}{F.~\surnamestart Mari\surnameend},
  \bibinfo{author}{I.~\surnamestart Melatti\surnameend},
  \bibinfo{author}{I.~\surnamestart Salvo\surnameend},
  \bibinfo{author}{E.~\surnamestart Tronci\surnameend}, \bibinfo{author}{J.~K.
  \surnamestart Gruber\surnameend}, \bibinfo{author}{B.~\surnamestart
  Hayes\surnameend}, \bibinfo{author}{M.~\surnamestart Prodanovic\surnameend}
  \& \bibinfo{author}{L.~\surnamestart Elmegaard\surnameend}
  (\bibinfo{year}{2014}): \emph{\bibinfo{title}{Demand-aware price policy
  synthesis and verification services for Smart Grids}}.
\newblock In: {\sl \bibinfo{booktitle}{2014 {IEEE} International Conference on
  Smart Grid Communications, SmartGridComm}},
  \doi{10.1109/SmartGridComm.2014.7007745}.

\bibitemdeclare{inproceedings}{iwbbio2015}
\bibitem{iwbbio2015}
\bibinfo{author}{T.~\surnamestart Mancini\surnameend},
  \bibinfo{author}{E.~\surnamestart Tronci\surnameend},
  \bibinfo{author}{I.~\surnamestart Salvo\surnameend},
  \bibinfo{author}{F.~\surnamestart Mari\surnameend},
  \bibinfo{author}{A.~\surnamestart Massini\surnameend} \&
  \bibinfo{author}{I.~\surnamestart Melatti\surnameend} (\bibinfo{year}{2015}):
  \emph{\bibinfo{title}{Computing Biological Model Parameters by Parallel
  Statistical Model Checking}}.
\newblock In: {\sl \bibinfo{booktitle}{Proc. Third Int. Conf. Bioinformatics
  and Biomedical Engineering, {IWBBIO}}}, \doi{10.1007/978-3-319-16480-9\_52}.

\bibitemdeclare{inproceedings}{MBR06}
\bibitem{MBR06}
\bibinfo{author}{B.~\surnamestart Meenakshi\surnameend},
  \bibinfo{author}{A.~\surnamestart Bhatnagar\surnameend} \&
  \bibinfo{author}{S.~\surnamestart Roy\surnameend} (\bibinfo{year}{2006}):
  \emph{\bibinfo{title}{Tool for Translating Simulink Models into Input
  Language of a Model Checker}}.
\newblock In: {\sl \bibinfo{booktitle}{Proc. 8th Int. Conf. on Formal
  Engineering Methods, ICFEM}}, \doi{10.1007/11901433\_33}.

\bibitemdeclare{inproceedings}{bcb2013}
\bibitem{bcb2013}
\bibinfo{author}{N.~\surnamestart Miskov{-}Zivanov\surnameend},
  \bibinfo{author}{P.~\surnamestart Zuliani\surnameend}, \bibinfo{author}{E.~M.
  \surnamestart Clarke\surnameend} \& \bibinfo{author}{J.~R. \surnamestart
  Faeder\surnameend} (\bibinfo{year}{2013}): \emph{\bibinfo{title}{Studies of
  biological networks with statistical model checking: application to immune
  system cells}}.
\newblock In: {\sl \bibinfo{booktitle}{{ACM} Conference on Bioinformatics,
  Computational Biology and Biomedical Informatics. {ACM-BCB}}},
  \doi{10.1145/2506583.2512390}.

\bibitemdeclare{inproceedings}{guiding-simulation-dac06}
\bibitem{guiding-simulation-dac06}
\bibinfo{author}{K.~\surnamestart Nanshi\surnameend} \&
  \bibinfo{author}{F.~\surnamestart Somenzi\surnameend} (\bibinfo{year}{2006}):
  \emph{\bibinfo{title}{Guiding simulation with increasingly refined abstract
  traces}}.
\newblock In: {\sl \bibinfo{booktitle}{Proc. 43rd annual Design Automation
  Conference}}, \bibinfo{series}{DAC '06}, \bibinfo{publisher}{ACM},
  \bibinfo{address}{New York, NY, USA}, \doi{10.1145/1146909.1147097}.

\bibitemdeclare{book}{sontag:book}
\bibitem{sontag:book}
\bibinfo{author}{E.D. \surnamestart Sontag\surnameend} (\bibinfo{year}{1998}):
  \emph{\bibinfo{title}{Mathematical Control Theory: Deterministic Finite
  Dimensional Systems}}.
\newblock \bibinfo{series}{Texts in Applied Mathematics},
  \bibinfo{publisher}{Springer}, \doi{10.1007/978-1-4612-0577-7}.

\bibitemdeclare{article}{TSCC05}
\bibitem{TSCC05}
\bibinfo{author}{S.~\surnamestart Tripakis\surnameend},
  \bibinfo{author}{C.~\surnamestart Sofronis\surnameend},
  \bibinfo{author}{P.~\surnamestart Caspi\surnameend} \&
  \bibinfo{author}{A.~\surnamestart Curic\surnameend} (\bibinfo{year}{2005}):
  \emph{\bibinfo{title}{Translating discrete-time simulink to lustre}}.
\newblock {\sl \bibinfo{journal}{ACM Trans. Embedded Comput. Syst.}}
  \bibinfo{volume}{4}(\bibinfo{number}{4}), \doi{10.1145/1113830.1113834}.

\bibitemdeclare{inproceedings}{fmcad2014}
\bibitem{fmcad2014}
\bibinfo{author}{E.~\surnamestart Tronci\surnameend},
  \bibinfo{author}{T.~\surnamestart Mancini\surnameend},
  \bibinfo{author}{I.~\surnamestart Salvo\surnameend},
  \bibinfo{author}{S.~\surnamestart Sinisi\surnameend},
  \bibinfo{author}{F.~\surnamestart Mari\surnameend},
  \bibinfo{author}{I.~\surnamestart Melatti\surnameend},
  \bibinfo{author}{A.~\surnamestart Massini\surnameend},
  \bibinfo{author}{F.~\surnamestart Davi\surnameend},
  \bibinfo{author}{T.~\surnamestart Dierkes\surnameend},
  \bibinfo{author}{R.~\surnamestart Ehrig\surnameend},
  \bibinfo{author}{S.~\surnamestart R{\"{o}}blitz\surnameend},
  \bibinfo{author}{B.~\surnamestart Leeners\surnameend},
  \bibinfo{author}{T.~H.~C. \surnamestart Kruger\surnameend},
  \bibinfo{author}{M.~\surnamestart Egli\surnameend} \&
  \bibinfo{author}{F.~\surnamestart Ille\surnameend} (\bibinfo{year}{2014}):
  \emph{\bibinfo{title}{Patient-specific models from inter-patient biological
  models and clinical records}}.
\newblock In: {\sl \bibinfo{booktitle}{Formal Methods in Computer-Aided Design,
  {FMCAD}}}, \doi{10.1109/FMCAD.2014.6987615}.

\bibitemdeclare{inproceedings}{VSDP12}
\bibitem{VSDP12}
\bibinfo{author}{R.~\surnamestart Venkatesh\surnameend},
  \bibinfo{author}{U.~\surnamestart Shrotri\surnameend},
  \bibinfo{author}{P.~\surnamestart Darke\surnameend} \&
  \bibinfo{author}{P.~\surnamestart Bokil\surnameend} (\bibinfo{year}{2012}):
  \emph{\bibinfo{title}{Test generation for large automotive models}}.
\newblock In: {\sl \bibinfo{booktitle}{IEEE Int. Conf. on Industrial Technology
  (ICIT)}}, \doi{10.1109/ICIT.2012.6210014}.

\bibitemdeclare{inproceedings}{WCMKS07}
\bibitem{WCMKS07}
\bibinfo{author}{M.~W. \surnamestart Whalen\surnameend}, \bibinfo{author}{D.~D.
  \surnamestart Cofer\surnameend}, \bibinfo{author}{S.~P. \surnamestart
  Miller\surnameend}, \bibinfo{author}{B.~H. \surnamestart Krogh\surnameend} \&
  \bibinfo{author}{W.~\surnamestart Storm\surnameend} (\bibinfo{year}{2007}):
  \emph{\bibinfo{title}{Integration of Formal Analysis into a Model-Based
  Software Development Process}}.
\newblock In: {\sl \bibinfo{booktitle}{Proc. 12th Int. Workshop Formal Methods
  for Industrial Critical Systems, FMICS}}, \doi{10.1007/978-3-540-79707-4\_7}.

\bibitemdeclare{inproceedings}{validation-with-guided-search-dac98}
\bibitem{validation-with-guided-search-dac98}
\bibinfo{author}{C.~H. \surnamestart Yang\surnameend} \& \bibinfo{author}{D.~L.
  \surnamestart Dill\surnameend} (\bibinfo{year}{1998}):
  \emph{\bibinfo{title}{Validation with guided search of the state space}}.
\newblock In: {\sl \bibinfo{booktitle}{Proc. 35th annual Design Automation
  Conference}}, \bibinfo{series}{DAC '98}, \bibinfo{publisher}{ACM},
  \bibinfo{address}{New York, NY, USA}, \doi{10.1145/277044.277201}.

\bibitemdeclare{inproceedings}{ClarkeHSCC10}
\bibitem{ClarkeHSCC10}
\bibinfo{author}{P.~\surnamestart Zuliani\surnameend},
  \bibinfo{author}{A.~\surnamestart Platzer\surnameend} \&
  \bibinfo{author}{E.~M. \surnamestart Clarke\surnameend}
  (\bibinfo{year}{2010}): \emph{\bibinfo{title}{Bayesian statistical model
  checking with application to Simulink/Stateflow verification}}.
\newblock In: {\sl \bibinfo{booktitle}{Proc. 13th ACM Int. Conf. on Hybrid
  Systems: Computation and Control, HSCC}}, \doi{10.1145/1755952.1755987}.

\bibitemdeclare{article}{fmsd2013}
\bibitem{fmsd2013}
\bibinfo{author}{P.~\surnamestart Zuliani\surnameend},
  \bibinfo{author}{A.~\surnamestart Platzer\surnameend} \&
  \bibinfo{author}{E.~M. \surnamestart Clarke\surnameend}
  (\bibinfo{year}{2013}): \emph{\bibinfo{title}{Bayesian statistical model
  checking with application to Stateflow/Simulink verification}}.
\newblock {\sl \bibinfo{journal}{Formal Methods in System Design}}
  \bibinfo{volume}{43}(\bibinfo{number}{2}), \doi{10.1007/s10703-013-0195-3}.

\end{thebibliography}

%

\end{document}